\documentclass[12pt]{article}
\usepackage{amsmath,amsfonts, amssymb,verbatim}
\newtheorem{pkt}{}[section]
\newcommand{\bpk}{\begin{pkt}\rm }
\newcommand{\epk}{\end{pkt}}

\newcommand{\QQ}{{^*\mathbb{Q}}}
\newcommand{\V}{{\mathbf V}}
\newcommand{\CC}{{^*\mathbb{C}}}

\newcommand{\inv}{^{-1}}

\newcommand{\U}{\mathbb{U}}

\newcommand{\uuu}{\mathfrak{u}}

\newcommand{\Oo}{\mathrm{O}}

\newcommand{\M}{{\bf M}}

\newcommand{\ii}{\mathbf{i}}

\newcommand{\PP}{{\bf P}}

\newcommand{\R}{{\mathbb R}}

\newcommand{\Q}{{\mathbb Q}}
\newcommand{\Z}{{\mathbb Z}}
\newcommand{\N}{{\mathbb N}}
\newcommand{\C}{{\mathbb C}}

\newcommand{\be}{\begin{equation}}
\newcommand{\ee}{\end{equation}}

\newcommand{\Ss}{\mathbb{S}}

\newcommand{\F}{\mathrm{F} }
\newcommand{\pr}{{\rm pr} }

\newcommand{\K}{{\rm K}}

\newcommand{\trd}{{\rm tr.d.}}

\newcommand{\NN}{\mathcal{N}}
\newcommand{\MM}{\mathcal{M}}

\newcommand{\E}{\mathrm{E}}

\newcommand{\pp}{\mathfrak{p}}
\newcommand{\lfr}{\mathfrak{l}}

\newcommand{\ZZ}{{^*\mathbb{Z}}}

\newcommand{\lm}{\mathsf{lm}}

\title{Structural approximation and
a Minkowski space-time lattice with Lorentzian invariance
}
\author{Boris Zilber
\\ University of Oxford\footnote{The data that supports the findings of this study are available within the article  } }

\begin{document}
\maketitle
\abstract{  
We introduce a framework of structural approximation to represent Lorentz-invariant Minkowski space-time as the limit of finite cyclic lattices, each equipped with the action of a finite quasi-Lorentz group. This construction provides a discrete model preserving Lorentz symmetry and offers new insights into the algebraic and geometric structure of space-time.}

\section{Introduction}
\bpk A physical theory is an approximation to reality. But what is an
approximation? In
\cite{Zilber2014} we discussed this problem from the perspective of model theory. One of the necessary properties of a physical theory is that the model preserves the same structural properties that is being assumed for reality.  Such properties  can often be formulated in formal languages studied by logicians and we used well-studied model-theoretic tools to introduce in  
\cite{Zilber2014}   the definition of {\bf structural approximation} and respective notion of limit, which
the current paper heavily relies upon. An earlier application of this notion in context of quantum and statistical mechanics is in
the recent paper \cite{Zilber2025}.   

Among the fundamental properties that physics may  ascribe to the  universe is the property of being symmetric with regard to the action of a certain group of transformations. On the other hand we might allow 
the possibility of the universe of being discrete or even finite (consisting of enormous number of points) while the present theory treats it as continuous.
 In such a setting structural approximation considers a sequence of
$\M_i,$ $i\in \N$, of finite structures each equipped with a finite group $G_i$ acting on $\M_i$ in a certain prescribed way. The definition
in section \ref{s2}, following \cite{Zilber2014}, makes precise the statement that $(\M_i,G_i),$ $i\in \N,$ {\bf approximates} a continuous model $\M$ with an action of continuous group $G$ along an ultrailter $\mathcal{D}$.   This can be written as
$$\lm_\mathcal{D} (\M_i,G_i)=(\M,G).$$

However, such a straightforward form of  structural approximation 
gives intuitively expected outcomes
%is applicable 
only in simple cases, basically when $G$ is abelian and compact. In more interesting cases, it turns out that both $\M$ and $G$ have to be both representable as {\em compact complex manifolds}, see \ref{AandC}. As a result $G$ is a {\em compactified group} of transformations, that is a group extended by {\em infinitary} transformations which send  elements of $\M$ to a connected components of its boundary (obtained as a result of approximation) rather than a single point. For this reason we call the approximation as defined in \cite{Zilber2014} {\bf global}.

This is, in effect, due to the fact that in the discrete setting the algebra of the meaningful $\M_i$ and $G_i$ is based on rings of the form $\Z/n_i\Z$ (residue ring modulo $n_i$) with $n_i\to \infty.$ It turns out that in the limit of structural approximation such rings have to be identified as the compactified field of complex numbers $\bar{\C}=\C\cup \{ \infty\}$ that is
$$\lm_\mathcal{D}\, \Z/n_i \Z= \bar{\C}$$
along any non-principal ultrafilter.
\epk
\bpk
There is another, more standard and intuitive form of approximation, which we refer to as {\bf local approximation}. This corresponds to ``an observer located inside $\M$ with a limited observation span''. Namely,  
suppose  a part of $\M_i$, has a form of a 1-dim lattice of a large size $l_i$ which still is much smaller than the size of $\M_i.$ It is convenient in this context to use the language and the setting of non-standard numbers (see e.g. \cite{Albeverio1986} for definitions and examples) and assume that we work in some pseudo-finite (equivalently, hyper-finite) structure ${^*\M}$ with a  lattice of pseudo-finite size $\lfr.$ 

Let us  call $\uuu$ the spacing  of the lattice.  
Set a non-standard  distance on the lattice so that the shift by  $k \cdot\uuu$ corresponds to the distance $\frac{k}{\lfr}.$ The latter
by definition is an element of the set $\QQ$ of non-standard rational numbers.
 Now use the standard part map (of non-standard analysis) $$\mathrm{st}: \QQ\to \R\cup \{+\infty, -\infty\}$$  into compactified reals to convert the interval $\{ k \uuu: \ 0\le k\le \lfr\}$ in the lattice into the interval $[0,1]\subset \R$ and more generally, the union of increasing intervals    
$$\mathrm{st}:\ \bigcup_{m\in \N}\{ k \uuu: \ -m\lfr\le k\le m\lfr\}\ \twoheadrightarrow\ \R$$
where $\R$ is the real line as a metric space. In case that the $\M_i$ have an algebraic structure of rings or  fields, as e.g. in the case of\linebreak $\Z/p_i\Z\, (=\F_{p_i})$ $p_i$ prime,  one can arrange that map $\mathrm{st}$ preserves  also relations of the structure, see \cite{Zilber2025},  Theorem 4.11.  The proof of this theorem in \cite{Zilber2025} furnishes a construction of $\lm_\mathcal{D}$ so that the limit map  coincides with the standard part map on the subset of non-standard rational numbers embedded in the pseudo-finite field. Thus this construction
  {\em locally approximates  $\R$ while globally approximates $\bar{\C}$ by a sequence of finite fields.}

\epk
\bpk
In the paper we apply these techniques and construct a sequence of
finite groups  acting on finite 4-dim lattices  which  {\bf locally approximates} the Lorentz group $\mathrm{SO}^+(1,3)$ acting on the Minkowski space-time $\mathcal{M}(\R)$ endowed with the Minkowski metric.

The same sequence of finite structures {\bf globally approximates} a compact complexification $\overline{\mathcal{M}}(\C)$ of Minkowski space  acted upon by a compactification of the group  $\overline{\mathrm{SO}}(4,\C).$ As it happens the structure is almost identical with Penrose's twistor space, see \cite{PenroseRindler1984}. The same approximation map considered {\bf locally}  induces a unique compactification $\overline{\mathcal{M}}(\R)$ of  $\mathcal{M}(\R)$ which turns out to be a conformal compactification. More precisely, 
$$\overline{\mathcal{M}}(\R)\subset \overline{\mathcal{M}}(\C)\hookrightarrow \C\PP^5,$$
with embedding into $\C\PP^5$ as the Grassmanian $\mathrm{Gr}(2,4),$ and
$$\overline{\mathcal{M}}(\R)\cong \R\PP^1\times \R\PP^3$$ 
(note that its double cover is  $\Ss^1\times \Ss^3$). 
 
Unlike Penrose's 
this compactification is truly cyclic: null rays are closed, in fact isomorphic to the projective line $\R\PP^1.$ This agrees well with the cyclic structure of the discrete lattices which structurally approximate Minkowski spacetime in our construction.
\epk
\bpk Finally we would like to mention without going here into details that the 4-dimensional lattice modelling Minkowski space and the structure around it fits well with constructions in \cite{Zilber2025}. Essentially, the 4-dimensional universe would be representable as $\U^4$ where $\U$ is the 1-dimensional universe of the earlier paper. 

The last section of the paper illustrates how $\U^4$ can be treated as a lattice model of Minkowski space-time endowed with solutions of Klein-Gordon equation  invariant under the  action of quasi-Lorentz group.
\epk
\section{Structural approximation by rings}\label{s2}
In this section we give precise definitions and quote main theorems  related to approximation, in particular, in the class of rings.    
\bpk {\bf Ultraproducts and pseudo-finite structures.}
A  structure $\M$ is a set $M$ with a collection $\Sigma$  of $n$-ary relations $S\subset M^n,$ for some $n,$ called the language (or the vocabulary) of $\M=(M;\Sigma).$  

Suppose we are given a sequence $\{\M_i: i\in \N\},$ $\M_i=(M_i,\Sigma)$, of structures in language $\Sigma.$ One can choose a Fr\'echet ultrafilter $D$ on $\N$ and construct the ultraproduct
$$  {^*\M}:=\prod_{i\in \N} \M_i/D$$
which is a structure in language $\Sigma$ with the
 key property  (the  Lo\'s theorem): given a first-order sentence $\sigma$ in the language $\Sigma,$ 
\be\label{Los} \sigma\mbox{ is true  in }{^*\M}\mbox{ if and only if }\sigma\mbox{ is true in }\M_i\mbox{ along }\mathcal{D} 
\ee
${^*\M}$ is often referred to as {\em the model-theoretic limit of $\M_i$} along $D$ (not a structural approximation yet).

In case  the $\M_i$'s are finite, ${^*\M}$ is said to be a {\bf pseudo-finite} structure (or hyper-finite in the terminology of \cite{Albeverio1986}).

In particular, if $\M_i=\F_{p_i},$ $p_i$ prime numbers, structure ${^*\M}$ is a pseudo-finite field of characteristic zero.
\epk
\bpk {\bf Structural approximation.}
It is convenient to consider the system $\mathcal{T}_n$ of  topologies on  ${\M}^n,$ all $n$, the basic closed sets of which are realisations $S(\M)\subset {\M}^n$ of the $n$-ary $S\in \Sigma$. 
In geometric/physics setting the closed sets could be introduced as the zero sets of appropriate systems of equations.  

Such a system is said to be quasi-compact (or complete) if the projection maps 
$\pr_{n+1,n}: {\M}^{n+1}\to {\M}^n$ preserve closed subsets, that is
$\pr_{n+1,n}(S)$  closed, for  $S$ closed. 

This definition makes sense for the $\M_i,$ $i\in I,$  and indeed for any structure in  language $\Sigma.$

A {\bf structural approximation} of $\M$ by $\{ \M_i: i\in I\}$ along $D$
is determined by a surjective map ({\bf limit map})  
\be\label{approx} \lm: {^*\M}\twoheadrightarrow \M\ee
which has the property, for any $n$-ary closed $S\in \Sigma$ and $s\in {^*\M}$:  
$$ s\in S({^*\M}) \mbox{ closed} \Rightarrow \lm(s)\in S(\M)$$
%where ``closed'' means  defined by positive quantifier-free formulas.

 As it happens, below, most of the time $\M_i,$ $\M$ and ${^*\M}$ are rings in the language $\{ x+y=z, \ x\cdot y=z\}$
 or groups in the language $x*y=z,$ and {\em closed } for  $S\subset \M^n$  means $S$ is the set of solutions of a system of algebraic equations in $n$-variables with parameters in $\M,$ or   equivalently,  closed in {\bf Zariski topology}.  
 
 Note that despite the coarse topology we still are able to use the intuition of infinitesimals:  two elements
$a,a'\in {^*\M}^n$ are seen to be ``infinitesimally close'' if
$\lm\, a=\lm\, a'.$

Thus, in view of (\ref{Los}) and (\ref{approx})  structural approximation is a formalisation of the statement {\em a very large structure $\M_i$ looks like $\M$ from afar.}  

\epk
\bpk\label{AandC} {\bf Approximation and compactness.}
It was established in \cite{Zilber2014} that in main cases  $\M$ has to be { quasi-compact} in the formal topology (complete in Zariski topology)
in order for it to be approximated by a non-trivial sequence $\M_i$ as in   (\ref{approx}).  %In particular,
%the field $\C$ is not quasi-compact but its compactification $ is.

For fields it takes the following form (Theorem 5.2 of \cite{Zilber2014}): 

{\em The compactification $\bar{\C}:=\C\cup\{ \infty\}=\C\PP^1$ of the field of complex numbers can be approximated by a sequence of finite fields $\F_{p_i}$ along an ultrafilter $D$ on prime numbers $p_i.$ 

Moreover, 
  for any  zero-characteristic  pseudo-finite field $\E$ 
  %(in particular, for $\E=\F_\pp,$ $\pp$ non-standard prime number)  
  there exists a structural approximation
\be\label{lmE}\lm_\E: \E\twoheadrightarrow \bar{\C}.\ee
%Warning: such an approximation {\em can not be explained in terms of non-standard analysis}. 

The field of complex numbers is the only locally compact field that can be structurally approximated  by a sequence of finite fields.}

\medskip

The limit map in (\ref{lmE}) is far of being unique but we can pick ones with some specific and useful properties as in \cite{Zilber2025}, which allows to mimic complex analysis in the pseudo-finite field $\E$.

\epk 
\bpk\label{2.4} {\bf Scales and scale-dependence of approximation.}

The interplay between the domain and the range of the approximation map $\lm_\E$ as in (\ref{lmE}) brings in some features not encountered in the limit construction with inherent metrics. By its nature  field $\E$ is of a pseudo-finite (non-standard) characteristic $\pp$ (``the limit'' of $p_i$) %(more generally we also consider pseudo-finite  residue rings $\E=\ZZ/\NN$ ) 
while $\C$ is characteristic zero field with a natural metric. 

More precisely, as in \cite{Zilber2025} let $\ZZ$ be an $\aleph_0$-saturated model of arithmetic, $\pp\in \ZZ$  an infinite prime number
 and
  $\E=\ZZ/\pp \ZZ.$

Note that in $\E$ usual integers $1,2,3\ldots$ are represented, and the sequence continuous so far as it does not reach the infinite pseudo-finite number $\pp,$ which is bigger than any standard number.
 
It is clear by algebraic considerations that $\lm_\E$ sends standard integers  $1,2,3\ldots$ of $\E$ to respective integers  $1,2,3\ldots$ of 
$\C.$ Moreover, on the domain of standard %(or more generally {\em small scale}) 
integers $\lm_\E$ acts as an isomorphism of subrings. The same is true  
for rational numbers, where we represent  rational numbers $\frac{k}{m}$ %(including non-standard) 
as pairs $(k,m)$ of integer numbers satisfying the identity 
$$(k_1,m_1)=(k_2,m_2) \Leftrightarrow k_1m_2=k_2m_1.$$
 
In \cite{Zilber2025}, 3.8-3.9 we defined a convex subring $\Oo$ (denoted 
$\Oo(\mathcal{F})$ therein) of the ring $\ZZ$ of {\bf small scale} non-standard integers. Namely,  
$$-\pp<< \Oo << \pp$$
and $\Oo$ is closed under applications of all arithmetic functions. In fact \be\label{Oo} \Oo\prec \ZZ,\ee  is an elementary  submodel of $\ZZ$ containing an infinite integers $\lfr$.   

Note that the field of fractions $\mathrm{Fr}(\Oo)$ of $\Oo$ can exactly be represented as 
$$\mathrm{Fr}(\Oo)=\{ (k,m): \ k,m\in \Oo,\ m\neq 0\}$$
and by construction \be\label{Fr}\mathrm{Fr}(\Oo)\subset \QQ,\ee the non-standard rational numbers, i.e. the fraction field of $\ZZ.$ 

We also use the field of {\em finite} rational numbers, that is
$$\mathrm{Fr}_\mathrm{fin}(\Oo)=\{ (k,m)\in \mathrm{Fr}(\Oo), \ \exists N\in \N:\ |k|< mN\}.$$
Note that by assumptions the natural map 
$$k\mapsto k\,\mathrm{mod}\, p; \ \Oo\to \E$$
is an embedding of rings, which extends to the embedding of fields, and we will assume for convenience of notation the inclusions\linebreak $\mathrm{Fr}_\mathrm{fin}(\Oo)\subseteq\mathrm{Fr}(\Oo)\subseteq \E$.

Define 
$$\E_\mathrm{Loc}:=\mathrm{Fr}_\mathrm{fin}(\Oo)$$

The construction in \cite{Zilber2025}, section 4, defines $\lm_\E$ on a subfield  $\mathrm{Fr}(\mathcal{F})$ of a pseudofinite field    to be {\bf the standard part map} $\mathrm{st},$ that is
\be \label{st} \lm_\E(x):=\mathrm{st}(x),\mbox{ for }x\in \E_\mathrm{Loc}\ee
while the values $\lm_\E(x)$ for  generic points of $\E$ have no explicit definition. 

Note that (\ref{st}) is well-defined because the standard part map is well-defined on $\QQ.$

We will call  an approximation (limit) map $\lm_\E$ satisfying (\ref{st}) as {\bf locally canonical}. In the paper all our limit maps are locally canonical by default. 
   
It is clear  that (\ref{st}) along with the fact that $\E_\mathrm{Loc}$ contains all finite rational numbers of a non-standard model of arithmetic implies 
\be\label{st-on}\lm_\E:\ \E_\mathrm{Loc}\twoheadrightarrow\R.\ee

In other words, an observer which has only access to small scale elements of $\E$ can think of $\E$ as being $\R.$ {\em Locally field $\E$ approximates the field of reals.}

However, as we continue along the natural  order $ 1,2,3,\ldots$
of $\E$, inevitably, (by  5.2(ii) of \cite{Zilber2014}) 
we will encounter elements of $\bar{\C}$ which are not in $\R,$ in particular an element $\ii\in \E$ such that $$\lm_\E: \ii\mapsto i=\sqrt{-1};\ \ \ \ii\cdot \E_\mathrm{Loc}\twoheadrightarrow i\cdot\R.$$ 

Again by (\ref{lmE}), we will also have a non-empty domain
$$\E_\infty=\{ x\in \E: \lm_\E(x)=\infty\}.$$
So, an observer which has tools to explore the global characteristics of $\E$ has to think of $\E$ as a Riemann sphere $\bar{\C}.$ 

{\em $\E$  locally looks like $\R$ while  globally looks like $\bar{\C}$}.

 \medspace
 
This, somewhat informal introduction of  local and global approximations is based on a mathematically  rigorous definition in \cite{Zlbeultra}. % by reduction to a  technical notion of {\em emergent metric ultraproduct}.  
\epk
\bpk\label{2.5} {\bf Approximation by rings.}  Consider a more general case of a ring of nonstandard integers $\ZZ$ modulo $\NN$  
$$\K=\ZZ/\NN \ZZ$$
where $\NN\in \ZZ$  divisible by all standard integers, the condition used in \cite{Zilber2025}.  (In the current setting it suffices to assume that $\NN$ is divisible by unboundedly large standard prime numbers).

\medskip

{\bf Lemma.} \label{2.5-} {\em There is a surjective ring-homomorphism 
$$\lm_\K:\ \K\to \bar{\C}.$$

%In particular,  $$\lm_\K (x+y)=
%\lm_\K\, x+ \lm_\K\, y,\mbox{ if \ } \lm\,x\neq\infty \mbox{ and \ } \lm\,y \neq\infty$$
%$$\lm_\K (x\cdot y)=
%\lm_\K\, x\cdot \lm_\K\, y,\mbox{ if \ } \lm_\K\,x\neq\infty \mbox{ and %\ } \lm_\K\,y \neq\infty$$

$\lm_\K$ is the composition of two Zariski homomorphisms
$$\mathrm{res}_\pp:\K \twoheadrightarrow \E\mbox{ and } \lm_\E: \E\twoheadrightarrow \bar{\C}$$
where $\E=\E_\pp=\ZZ/\pp \ZZ$ for some infinite prime,  a pseudo-finite field.
\medskip

There is an embedding $\mathrm{Fr}(\Oo)\subset \K$ such that
$$ \lm_\K(x):=\mathrm{st}(x)\mbox{ for }x\in \mathrm{Fr}_\mathrm{fin}(\Oo)$$ 
and, setting $\K_\mathrm{Loc}:=\mathrm{Fr}_\mathrm{fin}(\mathcal{F}),$
\be \label{Kloc}\lm_\K:\ \K_\mathrm{Loc}\twoheadrightarrow\R
\ee 
that is $\lm_\K$ is locally canonical.
}

{\bf Proof.} Since $\NN$, the order of $\K$, is divisible by every standard prime $p,$ there is a residue map, ring-homomorphism  $\mathrm{res}_\pp:\ \K\twoheadrightarrow\E_\pp,$ for an infinite non-standard $\pp.$ As in \ref{2.4} $\E_\pp=\E$ contains the non-standard rational numbers $\mathrm{Fr}(\Oo)$ and the subring $\E_\mathrm{Loc}$ which can be naturally lifted to $\K_\mathrm{Loc}.$

Now,  consider a large enough non-standard model $\CC$ of complex numbers so that it contains $\ZZ$ and thus embeds $\Oo,$ that is we may assume $\Oo\subset \CC.$ 
Since $\CC$ is algebraically closed of large enough transcendence degree
there is an embedding $$\mathrm{I}_\E: \E\hookrightarrow \CC$$
which is an identity on  $\mathrm{Fr}(\Oo).$  

Define $\lm_\E=\mathrm{st}\circ\mathrm{I},$   the composite of two maps.  This is a Zariski homomorphism.
Note that $\mathrm{st}(\mathrm{Fr}(\Oo))=\R\cup\{ \infty\},$ that is $\lm_\E(\E)\supseteq \R\cup \{ \infty\}.$ However  $\lm_\E(\E)\neq \R\cup \{ \infty\}$ by Theorem 5.2(ii) of \cite{Zilber2014}. It follows 
$\lm_\E(\E)=\bar{\C}.$

$\Box$

\epk
\bpk \label{Correction0} $$\K_\mathrm{Loc}:= \{ z\in \K: \mathrm{res}_\pp(z)\in \E_\mathrm{Loc}\}.$$
This is a subring which contains $\Oo$ and all inverses $z\inv$ modulo $\pp$ of elements $z\in \Oo.$
\epk
\bpk\label{lmV0} {\bf Approximation of algebraic varieties.} Let $V\subset \mathrm{A}^n$ be an affine variety over $\Z$ and $V(\K)$ its $\K$ points. By definition $V$ is given by a system of polynomial equations $P_V(x_1,\ldots,x_n)=0$ over $\Z$ and $$V(\K)=\{ (a_1,\ldots,a_n)\in \K^n: \ P_V(a_1,\ldots,a_n)=0\}.$$

We consider $V$ as a Zariski structure, that is a structure with universe $V(\K)$ and basic $k$-ary relations given as Zariski closed subsets $S\subseteq V(\K)^m$ defined over $\Q.$ 

Along with $V(\K)$ consider the complex variety $V(\C),$ also a Zariski structure by the same definition. There exists a projective variety $\PP$ over $\Q$ (not unique) such that  
$V$ can be embedded over $\Q$ into  $\PP$ as a quasi-projective subvariety. Let
$$e: V\hookrightarrow \PP$$
be the embedding.

Set $V^\PP\subseteq \PP$ be the Zariski closure of $e(V)$ in $\PP.$ It follows that $V^\PP(\C)$ is a complex projective variety, and so compact (complete). 
So $e (V(\R))\subset e (V(\C))\subset V^\PP(\C)$ and $e(V(\R))$ is a real algebraic variety isomorphic to $V(\R)$ over $\Q.$

Note that $V(\C)$ is a Zariski substructure of $V^\PP(\C)$ since every Zariski predicate $S\subset (V(\C))^m$ is a restriction of $S^\PP\subset(V^\PP(\C))^m,$ the closure of $S$ in $V^\PP(\C)$, to $V(\C).$

\epk
\bpk\label{lmV}
{\bf Theorem.} {\em Suppose $V(\Q)$ is dense in $V(\C)$ in the metric topology.
There exists a surjective Zariski homomorphism 
$$\lm_{V(\K)}: \ V(\K)\twoheadrightarrow V^\PP(\C)$$
%and 
%$$\lm_\K: \ \K \twoheadrightarrow \C\PP^1$$
%??? along with algebraic morphisms over $\Z$
%$$f: V^\PP(\C)\hookrightarrow (\C\PP^1)^n$$
such that the restriction of
$ \lm_{V(\K)}$ to $V(\K_\mathrm{Loc})$ is coordinate-wise the standard part map\footnote{Define  $\lm_{V(\K)}$ on 
$v\in V(\K_\mathrm{Loc})$ (defined in \ref{Correction0})
as $$\lm_{V(\K)}(v):=\mathrm{st}(\mathrm{res}_\pp(v)).$$}  
and
$$\lm_{V(\K)}: \ V(\K_\mathrm{Loc})\twoheadrightarrow e(V(\R))\subset V^\PP(\C).$$
}

{\bf Proof.} First consider the Zariski homomorphism
$$\mathrm{res}_\pp: \K\twoheadrightarrow \E;\ \ k\mapsto k\,\mbox{mod}\, \pp$$
for the prime $\pp$ and field $\E=\E_\pp$ described in \ref{2.5}.
%Recall that $\K_\mathrm{Loc}$ 

Extend $\mathrm{res}_\pp$ to $V(\K)$ coordinate-wise to get the Zariski homomorphism
$$\mathrm{res}_\pp: V(\K)\twoheadrightarrow V(\E).$$
This reduces the problem of constructing $\lm_{V(\K)}$ to the problem of constructing
$$\lm_{V(\E)}:\ V(\E)\twoheadrightarrow V^\PP(\C).$$
We carry out the construction below.

\medskip 

Define $$\lm_0: V(\E_\mathrm{Loc})\to V(\R); \ \ \lm_0(\bar{a})=\mathrm{st}(\bar{a}),$$  
where $\bar{a}=(a_1,\ldots,a_n)\in \E^n$ and $\mathrm{st}(\bar{a})=(\mathrm{st}(a_1),\ldots,\mathrm{st}(a_n)).$ 

By definition $\mathrm{st}$ is the map that preserves the order relation on rationals of $\E_\mathrm{Loc}$ and 
 preserves polynomial equations. It follows that $\lm_0$ is a Zariski homomorphism and tt is surjective due to (\ref{st-on}) and the assumption on density of rational points in $V(\R).$

At the next stage we require a slightly stronger version of (\ref{Oo}).

{\bf Claim.} {\em There exist a model  $\ZZ$ of the ring of integers, prime $\pp\in \ZZ$ and a convex elementary submodel $\Oo\prec \ZZ$ such that $\Z\neq \Oo$ and }
\be\label{trd}\trd(\E/\Oo)\ge 2^{\aleph_0}.\ee

Proof. First consider an $\aleph_0$-saturated model $\mathcal{Z}$ of the ring of integers  and note that $\Z\prec \mathcal{Z},$ by definition. It follows that there exists a prime $\pp\in \mathcal{Z}$ such that the field $\E_0=\mathcal{Z}/\pp \mathcal{Z}$ contains an infinite subset $A$
%$\{ \alpha_i: i< \aleph_0\}$
 of algebraically independent elements, that is $\trd(\E/\Z)\ge |A|\ge\aleph_0.$

Now let $D$ be a non-rpincipal ultrafilter over $\N$ and set the ultrapowers 
$$\Oo:= \Z^\N/D,\ \ \ZZ:= \mathcal{Z}^\N/D,\ \E:=\E_0^\N/D, \ {^*A}:= A^\N/D.$$
We then get $$\trd(\E/\Oo)\ge |{^*A}|\ge 2^{\aleph_0}$$
which proves the claim.

As a corollary we get: {\em There exists a subset of $n$-tuples $B\subset V(\E)\subset \E^n$ such that $|B|\ge  2^{\aleph_0}$ and  any $n$-tuples
$$(b_{11}\ldots b_{1n}),\ldots, (b_{m1}\ldots b_{mn})\in B$$ 
with distinct $b_{ij}$ are algebraically independent over $\E_\mathrm{Loc},$ that is
for any non-zero polynomial $P(x_{11}\ldots x_{1n},\ldots, x_{m1}\ldots x_{mn})$ over $\E_\mathrm{Loc},$ 
$$P(b_{11}\ldots b_{1n},\ldots, b_{m1}\ldots b_{mn})\neq 0.$$
}  

This implies that any map $\lm_B:\E_\mathrm{Loc}\cup B \to \V^\PP(\C)$ extending  $\lm_0$ is a Zariski homomorphism. We define the extension $\lm_B$ so that
\be\label{surj}\lm_B(B)= V^\PP(\C)\ee

%we aim at extending the partial homomorphism $\lm_0$ to a homomorphism $\lm_A$ defined on a subset $A\subset V(\E),$ $V(\E_\mathrm{Loc})\subset A,$ so that $\lm_A(A)=V^\PP(\C).$ 

It remains to extend $\lm_B$ to $A=V(\E)\setminus B.$ 

We start by enumerating $A$ by ordinal numbers $\alpha<\kappa$ where $\kappa=|A|,$ the cardinality of $A,$
$$A=\{ a_\alpha: \ a<\kappa\}.$$
We also extend the Zariski topologies on cartesian powers of $V(\E)$ and $V^\PP(\C)$ by adding, for every Zariski closed relation $S(x_1,\ldots,x_m)$  on $m$th cartesian power of the variety the new $k-1$-ary relation $\exists x_{m-k}\ldots x_m S(x_1,\ldots,x_m)$ (this procedure was called in \cite{Zilber2014} {\em the formal completion of the topology}). Note that each  relation $\exists x_m S(x_1,\ldots,x_m)$ on  
$V^\PP(\C)$ defines a  Zariski closed subset since $V^\PP(\C)$   is 
projective and thus complete.

Let $$A_\beta:=\{ a_\alpha\in A: \ \alpha<\beta\}.$$
Assume 
%by transfinite induction on 
that for $\beta<\kappa,$ there is 
$\lm_{B,\beta}:\ B\cup A_{\beta} \to  V^\PP(\C)$
such that, 
\be \label{beta}\begin{array}{lll} 
\mbox{for any }c_1,\ldots,c_{k-1}\in B\cup A_\beta,\\ 
 V(\E)\vDash \exists x_{k}\ldots x_{m} S(c_1,\ldots, c_{k-1}, x_{k} \ldots x_{m})\Rightarrow\\ 
\ \ \ \ \ V^\PP(\C)\vDash \exists x_{k}\ldots x_{m} S(c_1^\lm,\ldots,c_{k-1}^\lm,  x_{k}\ldots x_{m})\end{array}\ee
where $c^\lm$ stands for $\lm_{B,\beta}(c).$

This is the case for $\beta=0$ since 
$\lm_{B,0}:=\lm_B.$ 

Now we construct $\lm_{B,\beta+1}$ extending $\lm_{B,\beta}.$ We need to find an element $a_{\beta}^\lm\in V^\PP(\C)$ such that
$$\begin{array}{ll}  V(\E)\vDash \exists x_{k}\ldots x_{m-1} S(c_1,\ldots, c_{k-1}, x_{k} \ldots x_{m-1}, a_{\beta})\Rightarrow\\ 
\ \ \ \ \ V^\PP(\C)\vDash \exists x_{k}\ldots x_{m-1} S(c_1^\lm,\ldots,c_{k-1}^\lm,  x_{k}\ldots x_{m-1},  a_{\beta}^\lm)\end{array}$$
%where $c^\lm$ stands for (c).$

Consider the type over $A_\beta,$ collection  of  formulas with parameters $A_\beta,$ 
$$p(y)=\begin{array}{ll} \{ \exists x_{k}\ldots x_{m-1} S(c_1,\ldots, c_{k-1}, x_{k} \ldots x_{m-1}, y):\\  \ \ \ \ \ \ \ V(\E)\vDash \exists x_{k}\ldots x_{m-1} S(c_1,\ldots, c_{k-1}, x_{k} \ldots x_{m-1}, a_{\beta})\}\end{array}$$
We claim that $p(y)$ can be realised in $V^\PP(\C).$ 
%that is there is an element $
Indeed, each formula of $p(y)$ defines a Zariski closed subset of $V^\PP(\C)$ and the type corresponds to the intersection of all the Zariski closed subsets. By the descending chain condition of Zariski topologies $p(y)$ is equaivalent to the intersection of finitely many sets, equivalently to a single set defined by a formula  $\exists x_{k}\ldots x_{m-1} S(c_1,\ldots, c_{k-1}, x_{k} \ldots x_{m-1}, y).$
Thus we need to see that
$$V^\PP(\C)\vDash \exists y, x_{k},\ldots, x_{m-1}\, S(c_1,\ldots, c_{k-1}, x_{k} \ldots x_{m-1}, y)$$
which just follows from the indection assumption (\ref{beta}).

We have proved the induction step from $\lm_{B,\beta}$ to $\lm_{B,\beta+1}.$ For a limit ordinal $\gamma\le \kappa$ define $\lm_{B,\gamma}=\bigcup_{\beta<\gamma} \lm_{B,\beta}.$ This determines all the steps of the construction by induction and we set
$$\lm_{V(\E)}:= \lm_{B,\kappa}.$$
This finishes the proof of the theorem. $\Box$

\epk
\bpk \label{rational} {\bf Rational subvarieties.} Assume $Q\subset V$ a subvariety of dimension $d$ over $\Q$,  rational over $\Q.$ That is there are Zariski  open subsets $D\subseteq A^d,$ $Q^0\subseteq Q$ and  rational map $$f: D\twoheadrightarrow Q^0$$
all defined over $\Q.$ Let $Q^\PP$ be the Zariski closure of $Q$ in $\PP$ and assume that $Q^\PP(\C)$ is non-singular. 
 It is well-known that under the assumption $Q(\Q)$ is dense in $Q(\R)$ and in $Q^\PP(\R)$ in metric topology.

\medskip

{\bf Claim 1.} For  $\lm_{V(\E)}$ of \ref{lmV} 
 $$\lm_{V(\E)}(\E_\mathrm{Loc})=Q(\R)\mbox{ and }\lm_{V(\E)}(Q(\QQ))=Q^\PP(\R)$$
 where $\QQ$ is an $\aleph_0$-saturated model of the field of rationals embedded in $\E.$
 
 Proof. The first statement is a property of $\lm^0$ established in \ref{lmV}. 
 
 For the second statement consider $Q(\Q)$ which is dense in $Q(\R)$ and so dense in $Q^\PP(\R)$ since Zariski density implies metric density. 
Since the standard part map $\mathrm{st}$ sends a Cauchy sequence in $Q(\Q)$ to a Cauchy sequence in $Q^\PP(\R)$ and $Q(\QQ)$ is saturated, for each $a\in Q^\PP(\R)$ there is $\alpha\in Q(\QQ)$ such that
$\lm_{V(\E)}(\alpha)=a.$ This finishes the proof.

\medskip

{\bf Claim 2.} $$\lm_{V(\E)}(Q(\E))=Q^\PP(\C).$$

Proof. Recall that by \cite{Zilber2014} $\lm_{V(\E)}(\E)\neq \R\cup \{ \infty\}$ and so there  is $\gamma\in \E\setminus \QQ$ such that $\lm_{V(\E)}(\gamma)\in \C\setminus \R.$ Set $j:=\lm_{V(\E)}(\gamma).$
 Thus $\Q+j\cdot \Q$ is dense in $\bar{\C}$ in metric topology. 
 Note that since
$\lm_{V(\E)}$ is a homomorphism 
$$\lm_{V(\E)}(Q(\QQ+\gamma \cdot\QQ))= Q(\mathrm{st}(\QQ)+j\cdot \mathrm{st}(\QQ)).$$
Define an absolute value for $a+gb\in \Q+\gamma \Q\subset \E$ as
 $|a+gb|:=|a|+|b|,$ and similarly for $a+jb\in \R+j \R\subset \C$ as 
 $|a+jb|:=|a|+|b|.$ It follows that $\lm_{V(\E)}$ preserves the absolute value  and so
  sends Cauchy sequences 
 of  $\Q+\gamma \Q$ to Cauchy sequences 
 of $\Q+j\Q$. By the same argument as in the proof of Claim 1 we conclude that
$$Q(\mathrm{st}(\QQ)+j\cdot \mathrm{st}(\QQ))=Q^\PP(\C),$$
which completes the proof Claim 2. 

\epk
\bpk \label{Correction1} 
\epk

\section{Pseudo-finite Minkowski space structure and its limit}

%\bpk As in \cite{Zilber2025} let $\ZZ$ be an $\aleph_0$-saturated model of arithmetic, 
 %and
 % $\K=\K_\NN:=\ZZ/\NN$ be the (non-standard) residue ring.
 
%\epk
\bpk \label{X}  It is well-known that $\mathrm{SL}(2,\C)$ is isomorphic to the double cover of the 
Lorentz group $\mathrm{SO}^+(1,3)$ and it acts in agreement with this on  the Minkowski space .  

More precisely (see e.g. \cite{Klinker2015}),
one represents a vector with components $(t,x,y,z)\in \R^4$ (Minkowski space) as a $2 \times 2$ matrix
$$X:=\left(\begin{array}{ll}
t+z\ \ x-iy\\ 
 x+iy\ \ t-z \end{array}\right)$$

with $X^\dagger=X$ and $\det(X)=t^2-x^2-y^2-z^2,$ and considers
\be\label{LT} X\mapsto MXM^\dagger \mbox{ with } M\in \mathrm{SL}(2,\C).\ee
 This preserves $\det X$ and thus the Minkowski metric, which leads to the proof that (\ref{LT}) is  a Lorentz transformation and all   Lorentz transformations can be expressed in this way.
The fact that $\pm M$ both give the same transformation of $X$ corresponds to the fact that $\mathrm{SL}(2,\C)$ is the double cover of the Lorentz group, that is 
\be\label{O}\mathrm{SO}^+(1,3)\cong \mathrm{SL}(2,\C)/{\Z_2}\ee

We denote 
$$\left( \mathcal{M}(\R), \mathrm{SL}(2,\C)/{\Z_2}\right)$$
the structure which consists of $\R$-linear Minkowski space   $\mathcal{M}(\R)$ with metric given by $X\mapsto \det X$ along with the group   $\mathrm{SL}(2,\C)/{\Z_2}$ acting on the space as describe in (\ref{LT}). 

We note that the isomorphism of groups induces the isomorphism of structures
\be\label{iso}\left( \mathcal{M}(\R), \mathrm{SL}(2,\C)/{\Z_2}\right)\cong \left( \mathcal{M}(\R), \mathrm{SO}^+(1,3)\right)\ee

\epk

Let $\bar{\C}=\C\cup \{\infty\}$ which we will treat as a Zariski structure, that is the set with Zariski closed relations $R\subset \bar{\C}^n$ on it.

\bpk {\bf Complexification of a ring.}
Let $A$ be a commutative unitary ring. Define

$A^{(2)}$ to be the unitary ring obtained from the ring $A$ as follows:
$$A^{(2)}:= \{ (a,b)\in A\times A\};$$ $$(a_1,b_1)+(a_2,b_2):=(a_1+a_2,b_1+b_2),\ \ (a_1,b_1)\cdot(a_2,b_2):=(a_1a_2-b_1b_2, a_1b_2+a_2b_1).$$

Clearly, $a\mapsto (a,0)$ is an embedding of $A$ into $A^{(2)}$ as a subring $(A,0)$ and
$$(a,b)\mapsto (a,-b) \mbox{ an automorphism of }A^{(2)}.$$

\epk

\bpk\label{A-prop} Let $\mathrm{M}(2,A^{(2)})$ be the set of $2\times 2$ matrices over
$A^{(2)}$ which we treat as an $8$-dim $A$-module and let $\mathrm{SL}(2,A^{(2)})$  be the group of matrices of determinant 1.

A {\bf Minkowski $A$-lattice} is the $A$-submodule $\mathcal{M}(A)$ of $\mathrm{M}(2,A^{(2)})$
consisting of matrices $X_{t,x,y,z}$
 over $A^{(2)}$ of the form
$$X_{t,x,y,z}=X:=\left(\begin{array}{ll}
(t+z,0)\ \ \ \ (x,-y)\\ 
 (x,y)\ \ \ \ \ \ (t-z,0) \end{array}\right),\ \ \ t,x,y,z\in A.$$
 
We have  $$\det(X)=(t^2-x^2-y^2-z^2, 0)\in A\times \{ 0\}$$
and this defines Minkowski $A$-metric length of $(t,x,y,z).$

For the general  $A^{(2)}$-matrix
$$Y=\left(\begin{array}{ll}
(a_1,a_2)\ \ \ \ (b_1,b_2)\\ 
 (c_1,c_2)\ \ \ \  (d_1,d_2) \end{array}\right)$$
 define the adjoint matrix
 $$Y^\dagger:=\left(\begin{array}{ll}
(a_1,-a_2)\ \ \ \ (c_1,-c_2)\\ 
 (b_1,-b_2)\ \ \ \  (d_1,-d_2) \end{array}\right)$$

Clearly, $X^\dagger=X$ for $X\in \mathcal{M}(A).$  In general
$$(YZ)^\dagger=Z^\dagger Y^\dagger.$$
In particular, $Y$ is self-adjoint ($Y=Y^\dagger$) iff
$a_2=0=d_2$ and $b_1=c_1,$\linebreak $b_2=-c_2.$

It follows that for any   $M\in \mathrm{SL}(2,A^{(2)})$, $X \in \mathcal{M}(A)$ 
\be\label{LTK}  MXM^\dagger \in \mathcal{M}(A) \mbox{ and } \det X=\det MXM^\dagger \ee

Let $$C=\{ M\in \mathcal{M}(A):  
MXM^\dagger=X\mbox{ for all }X\in  \mathcal{M}(A).\}$$ 
Let $M_0\in C.$
In particular, $M_0M_0^\dagger=\mathrm{I}.$ It is equivalent to
$M_0^\dagger=M_0\inv$ and thus $M_0XM_0\inv=X$ for all $X\in \mathcal{M}(A).$ This  implies that $M_0$ is diagonal and belongs to  the centre of $ \mathrm{SL}(2,A^{(2)}),$ thus 
\be\label{C} C= \{ M=\left(\begin{array}{ll}
(a_1,a_2)\ \ \ \ 0\\ 
 0\ \ \ \  (a_1,a_2) \end{array}\right);\ \ \ a_1^2-a_2^2=1\ \& \ (a_1=0 \vee a_2=0)\}\ee

\medskip
 Thus we have established:
\epk
\bpk\label{3.4}
 {\bf Proposition.} {\em  The 2-sorted structure  $$\left(\mathcal{M}(A), \mathrm{SL}(2,A^{(2)})/C\right)$$ is interpretable in the ring $A$ along with the group action $X\mapsto  MXM^\dagger$ and  $A$-Minkowski metric.

The  action and Minkowski metric are defined by  systems of polynomial equations over $\Z$.
 
In particular,
$\mathrm{SL}(2,\K^{(2)})/C$  is the group of $K$-linear transformations of $\mathcal{M}(\K)$ preserving Minkowski $\K$-valued metric}.
 
\medskip 
 
%The fact that $\pm M$ both give the same transformation of $X$ corresponds to the fact that $\mathrm{SL}(2,\K^{(2)})$ is the cover of the $\K$-Lorentz group.
\epk

\bpk\label{SO} {\bf Lemma}. $$\mathrm{SL}(2,\C^{(2)})/C\cong \mathrm{SO}(4,\C)$$
where $C$ is the centre of  $\mathrm{SL}(2,\C^{(2)})$ and
$$C\cong \Z_2\times \Z_2.$$

{\bf Proof.} By \ref{3.4} $\mathrm{SL}(2,\C^{(2)})/C$ is the group of transformations of $\mathcal{M}(\C)$ preserving Minkowski $\C$-valued metric, that is the form $x_0^2+x_1^2+x_2^2+x_3^2.$ But this is also the definition of group $\mathrm{SO}(4,\C).$

The form of $C$ is determined by (\ref{C}).
 $\Box$
\epk
\bpk\label{3.8}  {\bf Compactification of Minkowski space}. 
Consider $\mathcal{M}(\C),$ the complexification of Minkowski space $\MM(\R).$ Clearly, setwise $\MM(\C)=\C^4.$ Note that the  metric in $\MM(\R)$ in presence of the additive structure  is determined by  the distance $s$ from $0,$ $s^2=-t^2+x_1^2+x_2^2+x_3^2,$
that is  a Zariski closed subset, quadric $Q(\R)\subset \MM(\R)\times \R.$ Similarly, $\mathrm{SL}(2,\R^{(2)})$ % $\mathrm{SO}^+(1,3)$ 
can be setwise  identified with a Zariski closed subset of $\R^8$ and the action of the group by a Zariski closed subset $\Gamma(\R)$ of
$\mathrm{SL}(2,\R^{(2)}) \times \MM(\R)\times \MM(\R)$. 

These all can be represented as a Zariski closed subsets  in cartesian powers $(\MM(\R)\times \R)^n,$ equivalently in $\R^{5n},$ for some $n.$

Let \be\label{VR} V(\R)= \mathcal{M}(\R)\times \R \times   
  \mathrm{SL}(2,\R^{(2)})\times \Gamma(\R)\ee
 the affine variety which represents in the form of Zariski closed sets the universes and the relation of the structure   $\left(\mathcal{M}(\R),
 \mathrm{SL}(2,\R^{(2)})\right),$ which is the structure of the Minkowski space-time with its metric given by $Q(\R)$ and the action of  the Lorentz group $\mathrm{SO}^+(1,3)$ represented by  $\mathrm{SL}(2,\R^{(2)})$ and $\Gamma(\R).$
 
 The same is applicable  to $\MM(\C),$ $\mathrm{SL}(2,\C^{(2)})$
and $\Gamma(\C),$
so we get $V(\C),$ the $\C$ points of the same variety, to represent the structure $\left(\mathcal{M}(\C),
 \mathrm{SL}(2,\C^{(2)})\right),$ the complexification of Minkowski structure.
  
\medskip

Following \ref{lmV}, for variety  $V$ choose an embedding into a projective variety $$e: V\to  \PP$$
(to be determined in the next section) and set $V^\PP$ to be Zariski closure of $e(V)$ in $\PP.$ We assume that $\PP$ is chosen so that the quadric $Q$ defining the Minkowski metric has a non-singular completion in $\PP.$

This gives us the $\PP$-compactification of the complex structure
$\left(\mathcal{M}(\C),
 \mathrm{SL}(2,\C^{(2)})\right)^\PP,$
 with compact ingredients
 $\overline{\MM}(\C),$ $\overline{\mathrm{SL}}(2,\C^{(2)})$ and  $\bar{\Gamma}(\C)$
as well as $\PP$-compactification of the real structure \linebreak
$\left(\mathcal{M}(\R),
 \mathrm{SL}(2,\R^{(2)})\right)^\PP$
 which will contain the compactification  $\overline{\MM}(\R)$ of Minkowski space.

\epk

\bpk\label{Thm} {\bf Theorem.} {\em There is a limit map, Zariski homomorphism of structures, 
\be\label{Lm0}\mathrm{Lm}: \left(\mathcal{M}(\K),
 \mathrm{SL}(2,\K^{(2)})\right)\twoheadrightarrow \left(\mathcal{M}(\C), \mathrm{SL}(2,\C^{(2)})\right)^\PP 
 \ee
 Its restriction  to the structure over
$\K_\mathrm{Loc}$ is a Zariski homomorphism

   \be\label{Lm1}
 \mathrm{Lm}:\left(\mathcal{M}(\K_\mathrm{Loc}),
 \mathrm{SL}(2,\K^{(2)}_\mathrm{Loc})\right)\twoheadrightarrow  
   \left(\mathcal{M}(\R), \mathrm{SL}(2,\C)\right)\ee

  }
 
 {\bf Proof.} Consider $V$ as defined in (\ref{VR}). 
 Note that  by the form of the factors of the direct product $\Q$-points are dense in each of the factors of $V(\R),$ or equivalently, 
 $V(\Q)$ is dense in $V(\R).$
 
  By  \ref{lmV} there is a Zariski homomorphism
  $$\lm_{V(\K)}: V(\K)\twoheadrightarrow V^\PP(\C)$$
  with the property
  $$\lm_{V(\K)}: V(\K_\mathrm{Loc})\twoheadrightarrow V(\R).$$
It remains to check that 
  $$\lm_{V(\K)}: Q(\K)\twoheadrightarrow Q^\PP(\C)$$
  and
  $$\lm_{V(\K)}: Q(\K_\mathrm{Loc})\twoheadrightarrow Q(\R).$$
  As in \ref{lmV} we can reduce the problem to proving
 $$\lm_{\V(\E)}: Q(\E)\twoheadrightarrow Q^\PP(\C)$$
  and
  $$\lm_{V(\E)}: Q(\E_\mathrm{Loc})\twoheadrightarrow Q(\R).$$ 
For this note that quadric $Q$  in the 5-dim affine space 
contains a $\Q$-point and
so is rational over $\Q.$   
Now \ref{rational} gives us both required properties.

 $\Box$ 
\epk

\section{Algebraic compactification of Minkowski space}
In this section we construct the complex projective variety $\PP$ in which
$\MM(\C)$ %and $\mathrm{SO}(4,\C)$ 
can be suitably embedded according to \ref{3.8}. Since we at the same time consider $\MM(\R),$ we carry out the construction over arbitrary field of characteristic $0$ to substitute $\C$ and $\R$ later.

The construction, which is usually applied to $\MM(\R),$ leading to a conformal compactification,
 is well known, although some small variations are possible. We found the algebraic presentation in \cite{Yadczyk1984}  useful for our purposes. The author is grateful to M.Kabenyuk for the help with the algebra involved. 
         
\bpk \label{8.2}
Let $\E$ be a field of characteristic $0,$
$\MM(\E)= \E \times \E^3$ be the Minkowski space with the quadratic form
encoding Minkowski metric
$$Q(t, x_1 , \ldots, x_3 ) = -t^2 + x^2_1 + \ldots + x^2_3.$$ 
 We define an embedding $\Phi$ of $\MM=\MM(\E)$ into a 
 higher-dimensional space $V = \E^2 \times \E^{4}$ as 
$$
\Phi:\ (t, x_1 , \ldots , x_3 ) \mapsto
(\frac{1}{2} -\frac{q}{2} , t, x_1 , . . . , x_3, \frac{1}{2} +\frac{q}{2})$$

where $q = Q(t, x_1 , . . . , x_3 ).$
The equation
$$\bar Q(s, t, x_1 , . . . , x_3 , x_{4} ) = -s^2 - t^2 +x^2_1+ , . . . , x^2_3 + x^2_{4}=0$$ 
defines a hypersurface $L$ in $V$. It is easy to verify that $\Phi(\MM ) \subseteq L.$

Let $\pi(V )$ be the projective space obtained by projectivisation 
$$\pi : V \to \pi(V ); \ (y_0,\ldots,y_{5})\mapsto [y_0:\ldots :y_{5}].$$
 We have dim $\dim \pi(V ) = 5,$ $ \dim \pi (L) = 4$ and
$$\pi (\MM ) \subset  \pi (L) \subset \pi (V ).$$

Moreover, $\pi$ bijectively maps $\Phi(\MM)$ onto $\pi (\Phi(\MM) ),$ (because $\Phi(\MM)$ lies in $s+x_{4}=1$)
so $\dim \pi (\Phi(\MM) ) = 4=\dim \pi(L).$

\epk
\bpk\label{8.3} Thus the map
$$\pi\circ\Phi:   (t, x_1 , \ldots , x_3 )\mapsto  [\frac{1}{2} -\frac{q}{2} : t: x_1 : . . . , x_3: \frac{1}{2} +\frac{q}{2}]$$ 
$$    \MM\twoheadrightarrow \pi(\Phi(\MM))\subset \pi(L)\subset \E\PP^{5}$$
is  bijective onto $\pi(\Phi(\MM))$ and $\pi(\Phi(\MM))$ is in the open subset of the projective variety $\pi(L)$ defined by equation $s+
x_{4}=1.$ 

Moreover, by the dimensional equalities $\pi(\Phi(\MM))$ is dense in $\pi(L).$ Thus we may identify the  closure in Zariski topology
$$\bar{\MM}:=\pi(L)=\{  [s : t: x_1 : . . . , x_3: x_{4}]\ : \ s^2 +t^2= x^2_1+ , . . . , x^2_3 + x^2_{4}=0\}$$ 
a smooth quadric in $\E\PP^{5}.$
 %and $P(\Phi(\MM))\subseteq P(L)$

\epk

\bpk\label{Gr} {\bf The complex case.} In case $\E=\C,$ the field of complex numbers, $\bar{\MM}(\C)$ by the above formula is a Klein quadric in $\C\PP^5$ and is easily identifiable with the complex Grassmanian $\mathrm{Gr}(2,4)\subset \C\PP^5.$

This identifies $\PP(\C)$ of \ref{Thm} as $\C\PP^5,$
\epk
\bpk {\bf The real case and conformal compactification.} In case $\E=\R$ it is possible to define the embedding 
 $$\psi: [s : t: x_1 : x_2: x_3: x_{4}] \mapsto  ([s : t], [x_1: x_2: x_3: x_{4}]);\ \ \pi(L)\to \R\PP^1\times \R\PP^{3}$$
 Indeed, given $(s, t, x_1, x_2, x_3, x_{4})\in L,$ non-zero,    we have 
\linebreak $s^2+t^2=x_1^2+ . . . + x_{4}^2.$ %and may assume $s+x_4=1.$ 
 Thus one of the affine coordinates $s,t$ has to be nonzero because otherwise $s^2+t^2=0=x_1^2+ . . . + x_{4}^2$ and so $s=t=x_1=x_2=x_3=x_4=0.$
For the same reason one of the coordinates of  $(x_1, x_2, x_3, x_{4})$ has to be non-zero. Thus 
$[s : t]\in \R\PP^1$ and $[x_1:x_2 : x_3: x_{4}]\in \R\PP^3$ which defines $\psi$. 

The embedding is  surjective. Indeed, let $[s : t]\in \R\PP^1$ and\linebreak $[x_1: . . . , x_3: x_{4}]\in \R\PP^3.$ Let $$\lambda=\sqrt{\frac{x_1^2+ . . . + x_{4}^2}{s^2+t^2}}$$ 
Then $[\lambda s: \lambda t]=[s:t]$ and $[\lambda s : \lambda t: x_1 : . . . , x_3: x_{4}]\in \pi(L).$ By definition  
 $$\psi:\ [\lambda s : \lambda t: x_1 : . . . , x_3: x_{4}]\mapsto  ([s : t], [x_1: . . . , x_3: x_{4}]).$$

 \medskip
 
 Now note that  $\R\PP^1\times \R\PP^3$ is homeomorphic to the quotient \linebreak $(\Ss^1\times\Ss^3)/\Z_2$ of products of the spheres, where $\Z_2$ acts diagonally. In fact the map 
  $$\psi':  \left(\frac{(s,t)}{\sqrt{s^2+t^2}}, \frac{(x_1,x_2,x_3,x_4)}{\sqrt{x_1^2+x_2^2+x_3^2+x_4^2}}\right)\mapsto  ([s : t], [x_1 : x_2: x_3: x_{4}])$$
  is a 2-1 map $\Ss^1\times\Ss^3\twoheadrightarrow \R\PP^1\times \R\PP^3.$ 
 
$\Ss^1\times\Ss^3$ is a well-known  conformal compactification of $\MM(\R).$ Since $\psi'$ preserves conformal metric, our algebraic  construction provides another known conformal compactification of $\MM(\R).$

 %$$s-x_4=s^2-x_4^2=-t^2+x_1^2+x_2^2+x_3^2=q$$ Under the assumption, 
% $[s : t]\mapsto (s,t)$ and  $ [x_1: . . . , x_3: x_{4}]\mapsto ( (x_1, %. . . , x_,: x_{d+1})$ is well-defined and so inverts $\psi.$ 
\epk

\bpk Set the  boundary of $\MM$ to be the difference
$$\mathcal{F}=\pi(L) \setminus \pi \Phi( \MM)=\bar{\MM}\setminus \MM.$$

The entire boundary is described by the single condition $s+x_4 = 0$ together
with the quadratic equation $\bar{Q}=0$ (as in \ref{8.2}). That cuts out exactly one quadric
in the hyperplane $\{ s + x_4 = 0\}$ in $\E\PP^5,$
 The hyperplane $\{ s +x_4 = 0\}$ is isomorphic to $\E\PP^4$ and does not intersect with $\MM$ (that is with $\pi \Phi( \MM)$).
  Restricting $\bar{Q}$ to this hyperplane yields a
nondegenerate  quadric 
$$Q^3: \ -t^2+x_1^2+x_2^2+x_3^2=0$$ of
 dimension 3 inside $\E\PP^4.$ Thus, $$\mathcal{F}\cong Q^3.$$
which, for $\E=\R$ is the {\bf light cone at infinity} in the terminology of R.Penrose.

\epk

\section{The Klein-Gordon equation as an example}
\bpk In \cite{Zilber2025} we considered a mathematical model $\U$ of a one-dimensional universe with wave-functions $\phi: \U^n\to \F_\pp.$  $\U$ is assumed to be a pseudo-finite residue group $\ZZ_\NN$ of  order $\NN$ such that $\NN$ is divisible by $\pp-1,$ for $\pp$ a  prime number, and $\F_\pp=\ZZ_\pp $ the pseudo-finite field with $\pp$ elements. Under the conditions there is a surjective homomorphism of groups
$$\exp_\pp: \U \twoheadrightarrow \F_\pp^\times$$

Now assume that $\U$ is the additive group of $\K=\K_\NN$ and $\NN.$ In particular $\U$ is a $\K$-module and we can identify $\U^4=\K^4$ with $\mathcal{M}(\K).$      
\epk
\bpk {\bf Klein-Gordon equation} %(Following \cite{MP}).

Recall the Klein-Gordon equation (with $\hbar=1=c$)
$$-\frac{\partial^2}{\partial t^2}\phi=(- \sum_{j=1,2,3} \frac{\partial^2}{\partial r^2_j}+m^2)\phi$$
which has solutions
$$\phi_\mathbf{k}  (\bar{r},t):= \exp(i \sum_{j=1,2,3} k_jr_j-i\omega t)$$ with \be \label{eqKG}\omega^2-k_1^2-k_2^2-k_3^2 =m^2\ee
%$$\phi(\bar{r},t):= \exp(i \bar{k}\cdot\bar{r}-i\omega t)$$
%where $$\bar{k}\cdot\bar{r}=\sum_{j=1}^{j=3}k_jr_j.$$

This can be rewritten in variables $\mathbf{x}=(x_0,x_1,x_2,x_3)$ and parameters $\mathbf{k}=(\omega,k_1,k_2,k_3)$
as 
\be\label{phiKG} \phi_\mathbf{k}  (\mathbf{x})=\exp(-i\, \mathbf{k}\cdot \mathbf{x}
)\ee
where $$\mathbf{k}\cdot \mathbf{x}=\omega x_0 - k_1 x_1-k_2x_2-k_3x_3$$
is the inner product in Minkowski space. It is well-known that by going through all  $\mathbf{k}\in \R^4$ satisfying (\ref{eqKG})  the solutions (\ref{phiKG}) generates the space of all solutions to the Klein-Gordon equation.  

Now we consider the analogue of solution (\ref{phiKG}) in $\mathcal{M}(\K)$ by assuming  $\mathbf{k},
\mathbf{x}\in \K^4$ and replacing $\exp (-ix)$ by $\exp_\pp(x).$
Thus (\ref{phiKG}) becomes 
\be\label{phiKK} \phi_\mathbf{k}  (\mathbf{x})=\exp_\pp(\mathbf{k}\cdot \mathbf{x}
)\ee

\epk
\bpk {\bf Lorentz invariance}: Recall that the substitute for the Lorentz group $\mathrm{SO}^+(1,3)=\mathrm{SL}(2,\C)/C$
is the $\K$-Lorentz group $\mathrm{SL}(2,\K^{2})/C$ as established in \ref{3.4}.

For $g\in \mathrm{SL}(2,\K^{2})$ consider the Lorentzian action of $g$ on $\mathbf{x}\in \mathcal{M}(\K)$
$$\mathbf{x}\mapsto \mathbf{x}^g$$
which transforms $\phi_\mathbf{k}  $:
$$\phi_\mathbf{k}  ^g(\mathbf{x}):= \phi_\mathbf{k}  (\mathbf{x}^g)$$

Note that 
\be\label{phig} \phi_\mathbf{k}  ^g(\mathbf{x})=\phi_{g\inv \mathbf{k}}(\mathbf{x})\ee
since $g$ preserves the metric on $\mathcal{\K},$ and so preserves Minkowski inner product, giving
$$\mathbf{k}\cdot \mathbf{x}^g= \mathbf{k}^{g\inv}\cdot \mathbf{x}$$

(\ref{phig}) proves that the action by $g$ on a solution produces just another solution, thus proving that the Klein-Gordon equation over $\mathcal{M}(\K)$ is invariant under $\K$-Lorentz action.
\epk

{\bf References}
\bibliographystyle{plainnat}

\bibitem[Z14]{Zilber2014} B.Zilber, {\em Perfect infinities and finite approximation}.  In: {\bf Infinity and Truth.}
IMS Lecture Notes Series, V.25, 2014 (also on author's web-page)
\bibitem[Z25]{Zilber2025} B.Zilber, {\em On the logical structure of physics and continuous model theory},  Monatshefte f\"ur Mathematik, May 2025
%\bibitem{CZ} TODD COCHRANE AND ZHIYONG ZHENG {\em A SURVEY ON PURE AND MIXED EXPONENTIAL SUMSMODULO PRIME POWERS }

%\bibitem[Hooft(2016)]{Hooft2016} G.t'Hooft, {\em How quantization of gravity leads to a discrete
%space-time}
 %2016 J. Phys.: Conf. Ser. 701 012014
%\bibitem{Etingof} P. Etingof, Geometry and Quantum Field Theory

\bibitem[Alb86]{Albeverio1986} S. Albeverio et al. {\bf Nonstandard Methods in Stochastic Analysis and Mathematical Physics}
%\bibitem{Palmer} T.Palmer, {\em The Invariant Set Postulate: a new geometric
%framework for the foundations of quantum
%theory and the role played by gravity}, Proc. R. Soc. A (2009) 465, 3165 - 3185
%\bibitem{EllisHawking} S.W. Hawking and G.F.R.Ellis, {\bf The Large Scale Structure of Space–Time}, CUP 1973
\bibitem[Z26]{Zlbeultra} B.Zilber, {\em  Approximation of structures: local and global} arXiv:2604.00720
\bibitem[Kl15]{Klinker2015} F.Klinker, {\em An explicit description of $\mathrm{SL}(2,\C)$ in terms of $\mathrm{SO}^+(3,1)$ and vice versa}, Intern. Electronc J. of Geometry, v.8, no 1, pp.94-104 (2015)

%\bibitem{AdamoTwistors}
%T. Adamo,  {\bf Lectures on twistor theory}, Proceedings of Science, vol. 323 (2018)
%\bibitem{FP} B.Zilber, {\em Physics over a finite field and Wick rotation},  arxiv: 2306.15698
\bibitem[Yadczyk84]{Yadczyk1984} Yadczyk, {\em On Conformal Infinity and Compactifications of
the Minkowski Space}, {\bf Advances in Applied Clifford Algebras}, 2010
\bibitem[PenRin84]{PenroseRindler1984} R. Penrose and W. Rindler, {\bf Spinors and Space-Time}, Vol. 2 – Spinor and
Twistor Methods in Space-Time Geometry, Cambridge University Press,
Cambridge, England, 1984
%\bibitem[Ax(1968)]{Ax1968} J.Ax, {\em The elementary theory of finite filds}, Annals 
%of Mathematics. Second Series, vol. 88 (1968),
%219 - 271 

\bibitem[MP]{MP} L.Salasnich, {\bf Modern Physics}, Springer 2022
\end{document}